\documentclass{article}

\usepackage{amsmath}
\usepackage{amssymb}
\usepackage{graphicx}
\usepackage{mhchem}
\usepackage{upgreek}
\usepackage{siunitx}
\usepackage{braket}
\usepackage[margin=1in]{geometry}
\usepackage{color}

\graphicspath{{figures/}}

\def\xdesign{\textit{XDesign}}

\def\nscan{n_{\textup{f}}}
\def\nscanos{n_{\textup{f,OS}}}
\def\nscanps{n_{\textup{f,PS}}}
\def\fprimeos{f^{\prime}_{\rm OS}}
\def\fprimeps{f^{\prime}_{\rm PS}}
\def\gammaos{\gamma_{\rm OS}}
\def\gammaps{\gamma_{\rm PS}}

\DeclareMathOperator*{\argmax}{argmax}

\title{X-ray tomography of extended objects: a comparison of data acquisition approaches}
\date{\vspace{-5ex}}

\begin{document}

\maketitle
\begin{center}
\small
Ming Du\textsuperscript{1},
Rafael Vescovi\textsuperscript{2,6},
Kamel Fezzaa\textsuperscript{2},
Chris Jacobsen\textsuperscript{2,3,4},
Do\u{g}a G\"{u}rsoy\textsuperscript{2,5,*}
\end{center}
\footnotesize
\noindent
\textsuperscript{1}{Department of Materials Science and Engineering, Northwestern University, Evanston, Illinois 60208, USA}\\
\textsuperscript{2}{Advanced Photon Source, Argonne National Laboratory, Argonne, Illinois 60439, USA}\\
\textsuperscript{3}{Department of Physics and Astronomy, Northwestern University, Evanston, Illinois 60208, USA}\\
\textsuperscript{4}{Chemistry of Life Processes Institute, Northwestern University, Evanston, Illinois 60208, USA}\\
\textsuperscript{5}{Department of Electrical Engineering and Computer Science,
  Northwestern University, Evanston, Illinois 60208, USA}\\
\textsuperscript{6}{Present address: Department of Neurobiology, University of Chicago, Chicago, Illinois 60637, USA}\\
\textsuperscript{*}{Corresponding author: dgursoy@aps.anl.gov}

\normalsize
\section*{Abstract}
  The penetration power of x-rays allows one to image large objects
  while their short wavelength allows for high spatial resolution.  As
  a result, with synchrotron sources one has the potential to obtain
  tomographic images of centimeter-sized specimens with sub-micrometer
  pixel sizes.  However, limitations on beam and detector size make it
  difficult to acquire such data of this sort in a single take,
  necessitating strategies for combining data from multiple
  regions. One strategy is to acquire a tiled set of local tomograms
  by rotating the specimen around each of the local tomogram center
  positions.  Another strategy, sinogram oriented acquisition,
  involves the collection of projections at multiple offset positions
  from the rotation axis followed by data merging and
  reconstruction. We have carried out a simulation study to compare
  these two approaches in terms of radiation dose applied to the
  specimen, and reconstructed image quality. Local tomography
  acquisition involves an easier data alignment problem, and immediate
  viewing of subregions before the entire dataset has been acquired.
  Sinogram oriented acquisition involves a more difficult data
  assembly and alignment procedure, and it is more sensitive to
  accumulative registration error.  However, sinogram oriented
  acquisition is more dose-efficient, it involves fewer translation
  motions of the object, and it avoids certain artifacts of local
  tomography.

\section{Introduction}

X-ray tomography offers a way to image the interior of extended
objects, and tomography at synchrotron light sources offers
significantly higher throughput than with laboratory sources when
working at $\sim 1$ micrometer voxel resolution or below.  However,
practical limitations of synchrotron x-ray beam width limit the size
of objects that can be studied in single field of view, and pixel
count in readily available image detectors sets a similar limit.  Thus
it becomes challenging to scale x-ray tomography up from the roughly
$(2000)^{3}=8$ gigavoxel volumes that are routinely imaged today,
towards the teravoxel volumes that are required for imaging
centimeter-sized objects at micrometer-scale voxel size.

\begin{figure}
  \centerline{\includegraphics[width=0.5\textwidth]{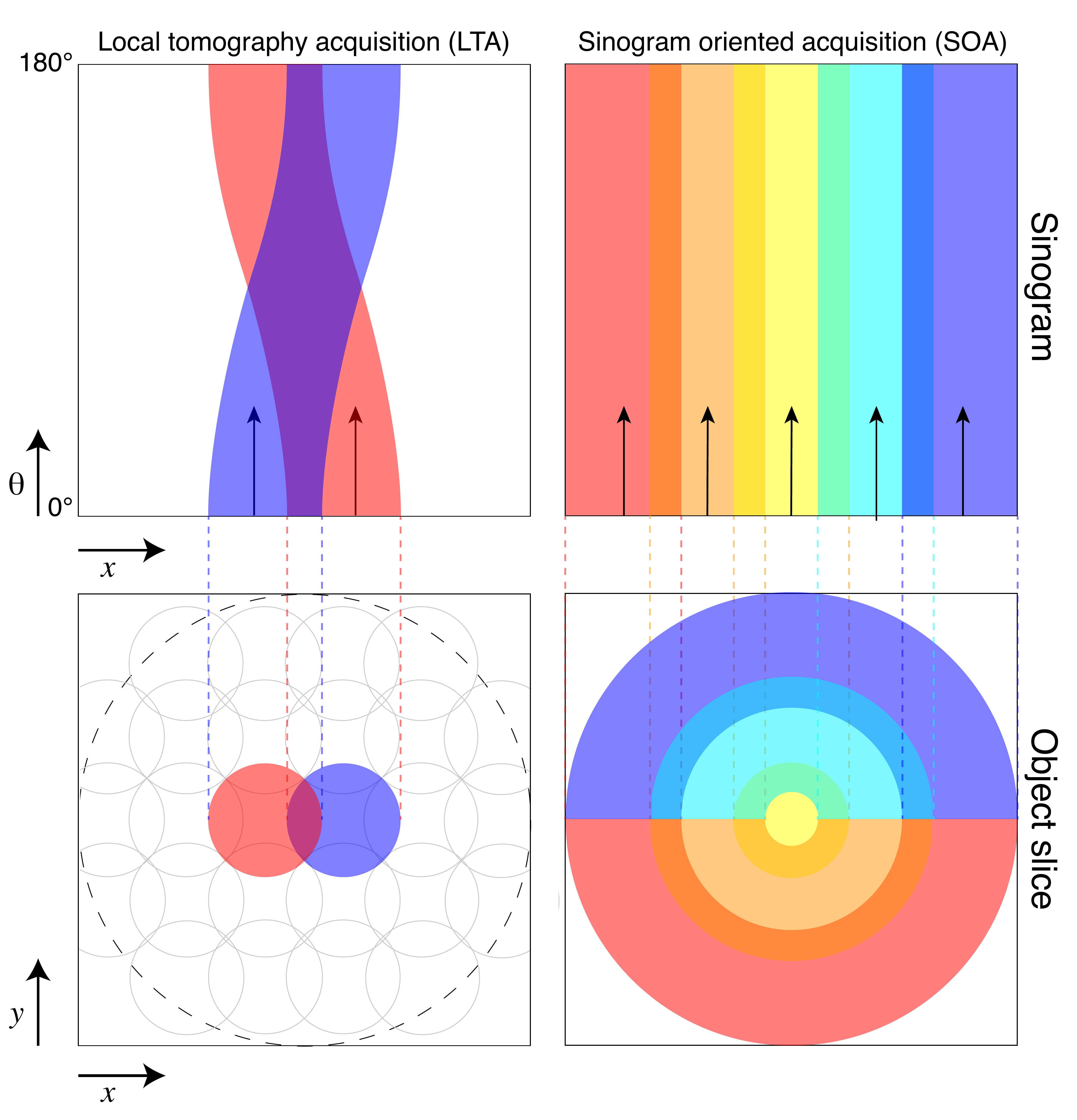}}
  \caption{Comparison on the acquisition scheme of local tomography
    acquisition (LTA; left) versus sinogram oriented acquisition (SOA;
    right). The top row depicts information collection in sinogram
    space, where each stripe with an arrow and a distinct color
    represents one angular scan over \ang{180} (which is then used to
    synthesize the full \ang{360} sinogram). The bottom row shows the
    mapping of different scans to the full image of one object
    slice. For samples with roughly equal extension in both lateral
    dimensions, if the number of scans required in SOA is $\nscan$,
    then $\nscan^2$ scans are needed by LTA.}
  \label{fig:acquisition}
\end{figure}

One solution lies in the use of one of several image stitching
approaches that can be applied to synchrotron x-ray tomography
\cite{kyrieleis_nima_2009}. Of those approaches discussed, we
consider here two of the most promising as shown schematically in
Fig.~\ref{fig:acquisition}:
\begin{itemize}

\item \textbf{Local tomography acquisition (LTA):} in
  local tomography \cite{kuchment_invprob_1995} (also called
  truncated object tomography \cite{lewitt_optik_1978a}, or interior
  tomography \cite{natterer_1986}), a subregion of a larger volume is
  imaged by rotating about the center of the subregion.  Features
  outside the subregion will contribute to some but not all
  projections, reducing their effects on the reconstructed image.  One
  can therefore acquire a tiled array of local tomograms to image the
  full specimen (method III of \cite{kyrieleis_nima_2009}).  In this
  case the rotation axis is shifted to be centered at each of the
  array of object positions, after which the object is rotated.  The
  local tomograms of the regions of interest (ROIs) are then
  reconstructed, and the full object is constructed from stitching
  together these local tomograms \cite{oikonomidis_jphysics_2017}.

\item \textbf{Sinogram oriented acquisition (SSA):} in this case, one
  acquires a set of ``ring in a cylinder'' projections
  \cite{vescovi_jsr_2017}.  The object is moved to a series of offset
  positions from the rotation axis, and at each position the object is
  rotated while projections are collected (method V of
  \cite{kyrieleis_nima_2009}).  The projections can be assembled and
  stitched to yield a full-field, panoramic projection image at each
  rotation angle, or they can be assembled and stitched in the
  sinogram representation. This method shares some common
  characteristics with the so-called ``half-acquisition'' method
  \cite{sztrokay_physmedbio_2012} in that both methods acquire
  sinograms of different parts of the sample, and stitch them before
  reconstruction (in half-acquisition, sinogram from \ang{180} to
  \ang{360} is flipped and stitched by the side of the \ang{0} to
  \ang{180} portion). The difference between them is that SOA can
  handle a larger number of fields in the horizontal axis (instead of
  2 in half-acquisition), and that each partial sinogram is acquired
  with the same rotation direction so no flipping is needed.

\end{itemize}
Another approach that has been employed with much
success involves
collecting a mosaic array of projection images at each rotation angle
\cite{liu_jsr_2012,mokso_jsr_2012} before repeating the process at the
next rotation. For each angle, the projections are assembled and
stitched to yield a full-field panoramic projection.  However, since
in practice it is usually quicker to rotate the specimen through
\ang{180} than it is to translate to a new mosaic offset position,
this approach (method I of \cite{kyrieleis_nima_2009}) has lower
throughput so we do not consider it further. Other large-scale imaging
methods like helical tomography
\cite{kalender_sucm_1994,pelt_measscitech_2018} are not discussed
here, as they have not been implemented for sub-micrometer resolution
imaging.  Therefore, we limit our discussion to LTA versus SOA as
defined above.

LTA and SOA are two distinct data collection strategies, each with
their own tradeoffs. For example, in LTA one can begin to reconstruct
regions of the object immediately after collection of its local
tomography data, whereas in SOA one must wait for the collection of all
``ring in a cylinder'' data before obtaining a full volume
reconstruction. One study of LTA
\cite{kyrieleis_jmicro_2010} indicated that the method contains
inherent complicating factors that can affect image quality, while
another study \cite{dasilva_optexp_2018} has shown that the
tomographic reconstruction of a local region can be improved
by using a multiscale acquisition approach including lower resolution
views of the entire specimen (this is not straightforward
when the specimen is larger than the illuminating beam).  However, we
are not aware of detailed comparisons of LTA and SOA 
with regards to radiation dose
efficiency as well as
reconstruction quality. Low radiation dose is critical
for X-ray imaging of soft materials, since they are
vulnerable to beam-induced damage and distortion
\cite{reimer_scanning_1984}.  Moreover, other factors may also come into
play when one does either SOA or LTA in practice. For example,
mechanical instabilities in translational motors introduce positional
fluctuations of the collected field-of-views, which requires image
registration to refine the relative offsets between them. For that,
SOA and LTA data behave differently in the presence of noise. Thus, a
comprehensive comparison is made here.

\section{Methodology}

\subsection{Phantom object}

\begin{figure}
	\centerline{\includegraphics[width=0.5\textwidth]{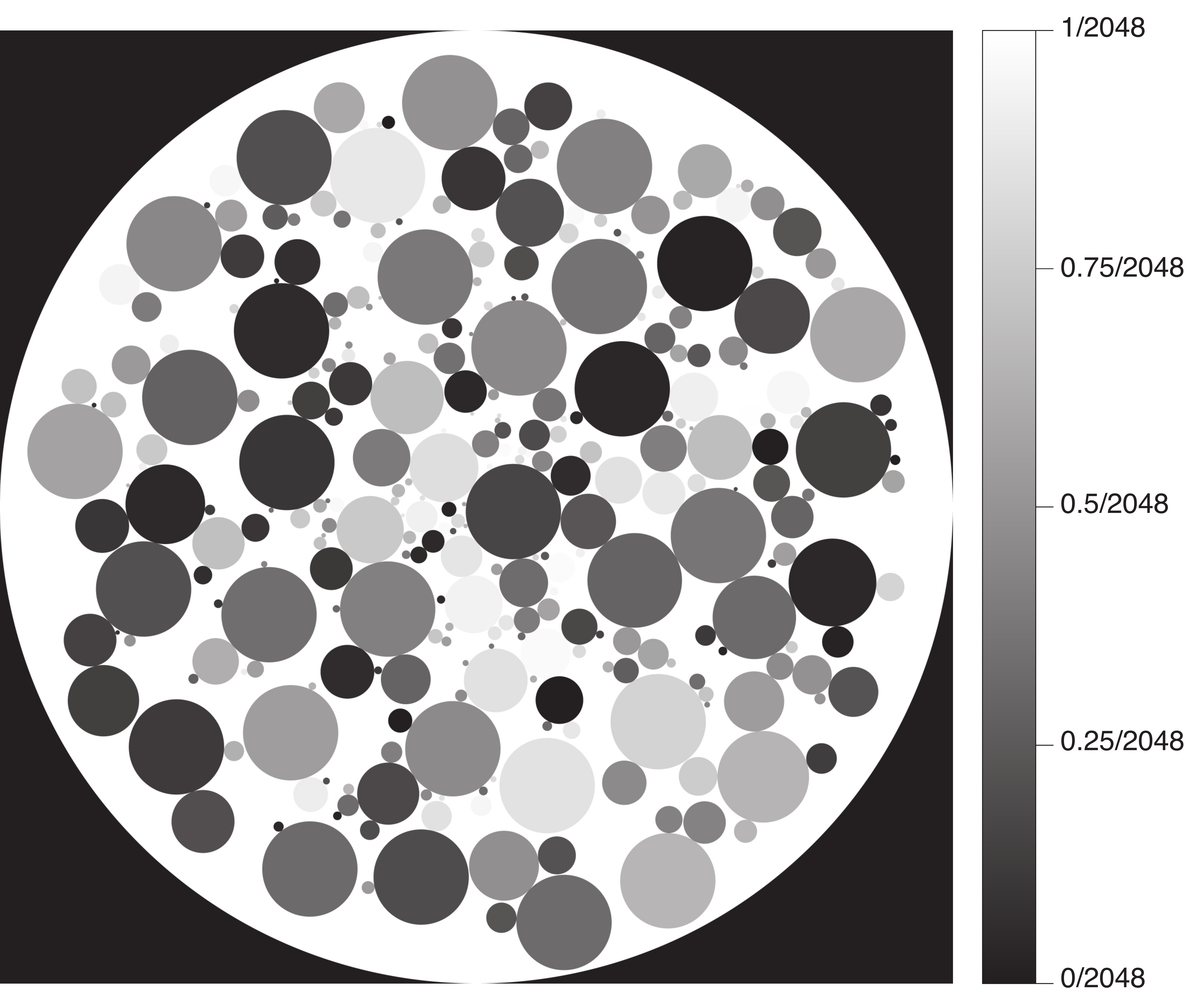}}
	\caption{Phantom object created for simulations, where the
          highest attenuation values are white and the lowest are
          black.  The object has a diameter, or maximum projected
          thickness, of $L=2048$ voxels, each set to a per-voxel
          linear attenuation coefficient of $\mu=1/2048$ so that the
          total attenuation through the disk if solid is $\exp[-1]$.
          The object is designed with random spherical
          ``pores'' inside with linear attenuation coefficients
          ranging from $\mu=0/2048$ to $\mu=1/2048$.}
	\label{fig:simulated_sample}
\end{figure}

In order to better understand the tradeoffs between object and
sinogram oriented acquisition, we created a 2D phantom sample using the
open-source virtual object designing tool \xdesign{}
\cite{ching_jsr_2017}. This represents an object slice from a 3D
object.  The simulated sample (Fig.~\ref{fig:simulated_sample}) is a
solid disk with a diameter, and thus maximum projected thickness, of
$L=2048$ pixels.  If solid, each pixel would be set to a linear
absorption coefficient (LAC) of $\mu=1/2048$, so that its total
thickness of $L=2048$ pixels would attenuate the x-ray beam by a
factor of $\exp[-\mu L]=\exp[-1]$.  In fact, the object was created
with circular pores in its interior, with diameters ranging from 8 to
205 pixels, and linear absorption coefficients ranging from
$\mu=0/2048$ (vacuum) to $\mu=1/2048$ (solid). All pores are randomly
distributed with no overlapping.  The object is also assumed to be
fully within the depth of focus of the imaging system, with no wave
propagation effects visible at the limit of spatial resolution, so
that pure projection images are obtained.  
%This also means that
%one can synthesize the full \ang{360} sinogram from data taken over a
%rotation range of only \ang{180}, as shown in Fig.~\ref{fig:acquisition}.
To generate the sinogram of the object, the Radon transform was
performed using \textit{TomoPy}, an open-source toolkit for x-ray
tomography \cite{gursoy_jsr_2014}.  All tomographic reconstructions in
this work are also obtained using this software package.

\subsection{Sampling for LTA and SOA}

%The sinogram of the object is
%generated in the assumption that the pure-projection criterion is
%valid. The partial sinograms acquired in SOA and LTA are obtained by
%sampling the full sinogram in the corresponding manner. The sampling
%scheme for sinogram oriented acquisition (SOA) is straightforward, as the coverage of the individual
%tile scans are just rectangular regions in the full sinogram, with the
%field-of-view (FOV) size $f$ as its width and the number of projection
%angles $N_{\theta}$ as its height. The horizontal (in the spatial axis) extent of
%the rectangle exactly matches the coverage of the FOV on the
%ray-projection of the sample.

\begin{figure}
	\centerline{\includegraphics[width=0.8\textwidth]{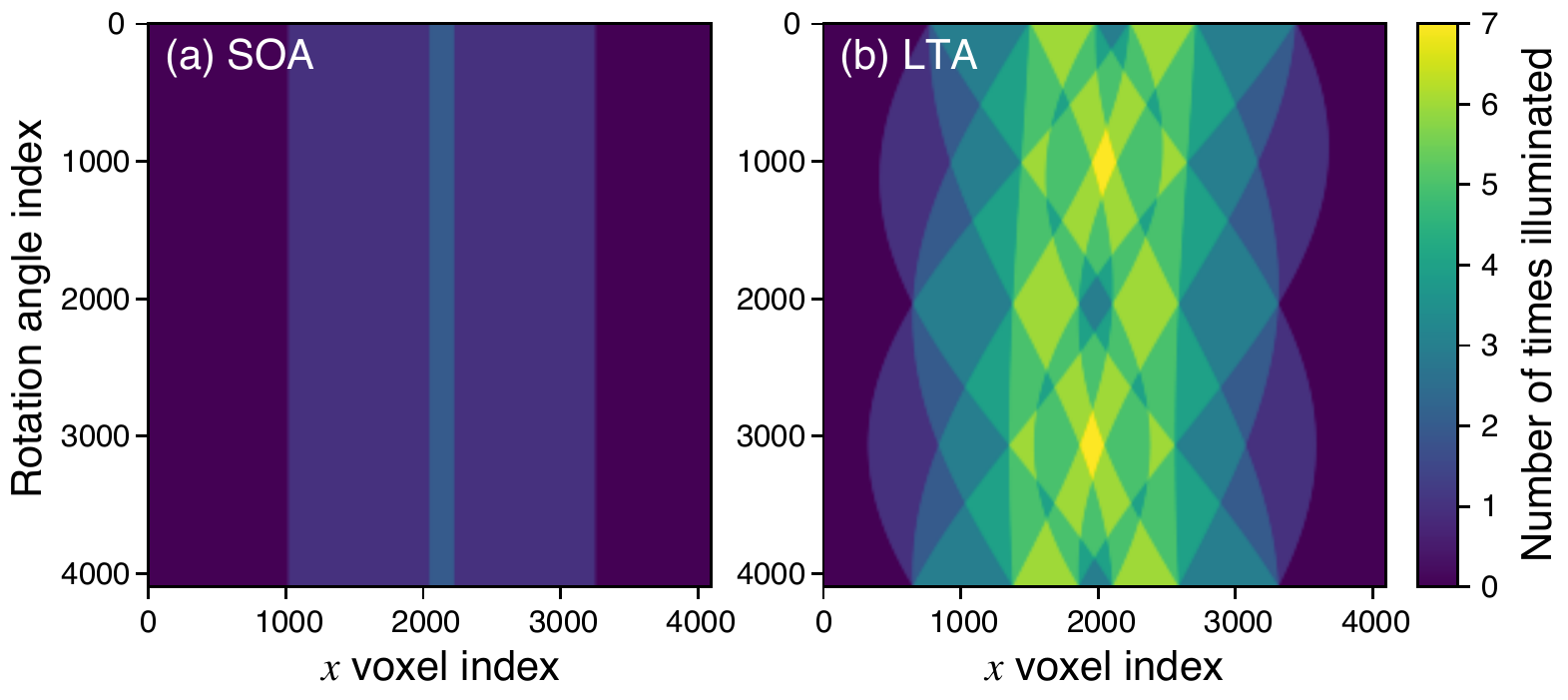}}
	\caption{Coverage on the full sinogram in an experiment using
          (a) sinogram oriented acquisition (SOA) and (b) local
          tomography acquisition (LTA) with equal field-of-view.
          Brighter values in the images correspond to the number of
          times that avoxel in the object is sampled by (exposed to) the
          illuminating beam.}
	\label{fig:sampling_map}
\end{figure}

To image an object larger than the imaging system's field of view
$f$, one provides some overlap between acquired projection
scans. The acquisition scheme can be conveniently 
shown in the sinogram
domain which contains both a spatial dimension and a viewing angle
dimension. A scan can be represented by a band-shaped coverage on
the sinogram, which is the region where we have access to the
measurement. Figure~\ref{fig:sampling_map} illustrates this coverage
for SOA and LTA, respectively, with the same field-of-view size for
both schemes. For LTA, a 3$\times$3 square grid is
used. Brighter values in the images means that a pixel in the sinogram
is sampled by the illuminating beam more frequently.

For LTA, by defining a coordination system with the 
origin $(0,0)$ located at the object center,
it can be shown that the coverage of a local tomography scan centered
at $(x, y)$ is a set of points on the \ang{360} synthesized sinogram given by
\begin{equation}
	C_{\textup{LTA}} = \{(s, \theta)|s_0(\theta) - f/2 \leq s \leq
        s_{0}(\theta) + f/2\}
        \label{eqn:c_os}
\end{equation}
with
\begin{equation}
	s_{0}(\theta) = \sqrt{x^2+y^2}\,\cos(\alpha-\theta) + c_{0}
        \label{eqn:s_0_theta}
\end{equation}
where $s$ and $\theta$ are respectively the horizontal (spatial) and
vertical (angular) coordinates of the sinogram, $\alpha = \arctan(x/y)$, 
and $c_0$ is the rotation center of the
synthesized \ang{360} sinogram.  This represents a partial sinogram of
the entire object, as shown in Fig.~\ref{fig:acquisition}. 
For SOA, the coverage is simply a straight band extending through
the angular axis. Mathematically, it can be expressed as
\begin{equation}
	C_{\textup{SOA}} = \{(s, \theta)|p - f/2 \leq s \leq p + f/2\}
        \label{eqn:c_ps}
\end{equation}
where $p$ is the center position of the field-of-view.

%First, each row in the sinogram is weighted such that the
%left and right boundaries are set to 1 (that is, the normalized
%transmitted intensity corresponding to zero absorption).  This is
%essential to prevent periodicity-related artifacts. Next, 
For object
stitching (LTA), the partial sinograms are padded with their edge 
values for twice their width on both sides
in order to reduce boundary artifacts in the reconstruction images
\cite{kyrieleis_jmicro_2010}. After reconstructing all partial
sinograms, the reconstructed disks are then stitched together to form
the complete reconstruction. 

%\textcolor{blue}{might be removed: In our acutual implementation, we only
%take the content within a fraction $\gammaos=0.85$ of the
%reconstruction disk radius, and discard the outer portion that is
%subject to intensity gradient artifacts near the boundary. Depending
%on the number of tiled scans $\nscanos$ along one edge of the
%reconstructed volume of an object of diameter $D$, this yields a
%reduced field of view $\fprimeos$ of
%\begin{equation}
%  T = \frac{f}{D}
%  \label{eqn:truncation_ratio}
%\end{equation}
%where $f$ is the FOV size over which data is collected (that is, the
%field of view of the camera), and $D$ is the actual sample size as
%before. The truncation ratio is closely related to the number of scans
%needed to image the entire sample. We define a quantity $\nscan$ that
%represents the number of scans along one side of the object that is
%required to fully reconstruct one slice of the sample.
%For SOA, this is equal to the total number of scans; for LTA, the total
%number of scans is $\nscan^2$ considering a square field
%to be stitched, though the actual number can vary depending on the object
%shape.}

Since both SOA and LTA involve multiple scans, we define a quantity
$\nscan$ that represents the number of scans along one side of the
object that is required to fully reconstruct one slice of the sample.
For SOA, $\nscanps$ is equal to the total number of scans; for LTA,
the total number of scans is roughly $\nscanos^2$ considering a square
grid of regions of interest (ROIs), though the actual number can vary
depending on the object shape. For example, applying LTA on a thin
sheet-like sample only requires roughly the same number of scans as
SOA. Also, one could choose hexagonal grids which are more efficient
by a factor of $\sqrt{3}$ than a square grid
\cite{heinzer_neuroimg_2006}, but we assume square grids here for
simplicity.

In order to fully reconstruct one slice of the object using SOA, there
should be a sufficient number $\nscanps$ of scan fields to guarantee
that the composite field of view completely covers the longest lateral
projection of that slice. In practice, an overlap that takes a
fraction $\gammaps$ of the field of view between each pair of adjacent
scans is needed for an automated algorithm to determine the offset
between them. With this taken into account, $\nscanps$ can be denoted
as
\begin{equation}
	\nscanps = \mbox{ceil}\Big[\frac{L - f}{\gammaps f} + 1\Big]
\end{equation}
where the function $\mbox{ceil}(x)$ is the ceiling function that returns the
smallest integer that is greater than or equal to a real number $x$. Since the overlapping area
diminishes the actual sample area that a scan can cover, we introduce a
``useful field of view'' $\fprimeps$ for SOA, given by $\fprimeps = \gammaps f$. 
For example, if a 15\% overlap is deliberately created between
a pair of adjacent scans, then $\fprimeps$ will be 85\% of the instrumental field-of-view.
Unless otherwise noted, in
this work we keep the value of $\gammaps$ to be 0.85 for simulation studies. 

\begin{figure}
	\centerline{\includegraphics[width=0.5\linewidth]{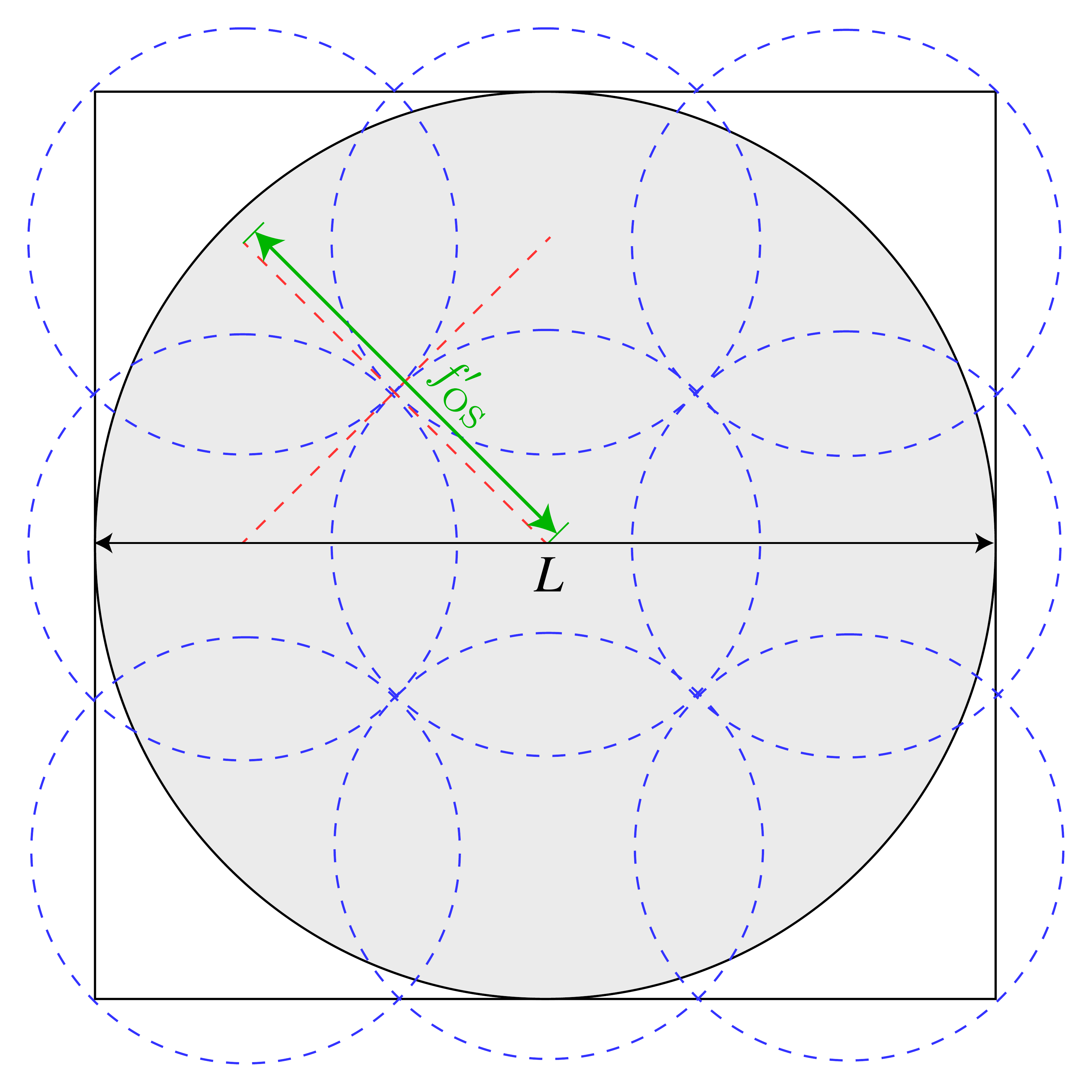}}
	\caption{Schematic diagram showing the assumed pattern of data
	acquisition in the LTA approach of beyond-field-of-view tomography. The
	specimen is represented by the gray solid disk. Each local ROI
	that can be reconstructed using the data acquired from a scan
	in LTA is shown by a dashed blue circle. Each of these ROIs has
	a diameter of $\fprimeos$, and they are packed in a way that the
	distance between the centers of each pair of diagonally adjacent
	ROIs is exactly $\fprimeos$, so that the sample can be fully
	covered without gaps. }
	\label{fig:os_sampling_scheme}
\end{figure}

The case for LTA differs in that the scans need to cover the object slice
in two dimensions. In principle, the scans in LTA can be arranged in an
arbitrary pattern that complies with the actual shape of the sample. If the
sample is square, then a roughly equal number of scans $\nscanos$ is needed
along both sides of the object, or $\nscanos^{2}$ scans in total. A special
notice should be paid to the width of the field of view in LTA, as it might not be equal to 
the actual
field of view of the optical system. In LTA, it has been found that 
the reconstructed ROI often exhibits a bowl-shaped
intensity profile, with values of near-boundary pixels abnormally
higher \cite{kyrieleis_jmicro_2010}. Although this can be mitigated
by padding the partial sinograms, this remedy does not work effectively
when the truncation ratio is very low. In such scenario, the 
reconstructed ROIs need to have a portion of their outer pixels
removed before they can be stitched. Similar to the case of SOA, 
we therefore introduce a ``useful field of view'' $\fprimeos$ for LTA.
If we use for local tomography acquisition only the content within
a disk whose radius is a fraction $\gammaos$ of the original ROI, 
then $\fprimeos = \gammaos f$. 
Consequently, the required number of scans $\nscanos$ is given by
\begin{equation}
    \nscanos = 
    \begin{cases}
    	1 & f \geq L \\
    	\mbox{ceil}\Big(\frac{\sqrt{2}L}{\gammaos f}\Big) & f < L
    \end{cases}.
    \label{eqn:nscan_os}
\end{equation}
We emphasize that $\nscanos$ is the number of scans required
along one side of the sample; for a square specimen, the total number
of scans needed is $\nscanos^2$. 
Eq.~\ref{eqn:nscan_os} is derived assuming the scenario indicated in
Fig.~\ref{fig:os_sampling_scheme}. When $f < L$, scanned
ROIs are arranged in a square grid such that each corner of the bounding
square of the sample disk intersects with the border of an ROI.
Also, we assume that  
the distance between the centers of two diagonally overlapping
ROIs is $\fprimeos$ so that all ROIs exactly cover the object seamlessly. 
Unless specifically indicated, the value of
$\gammaos$ is chosen to be 0.85 for simulation studies involved in this work. 

In order to understand the consequences of different object diameters
$L$, we follow previous work \cite{kyrieleis_jmicro_2010} and
characterize them in terms of a truncation ratio $T$ of
\begin{equation}
  T = \frac{f^{\prime}}{L}
  \label{eqn:truncation_ratio}
\end{equation}
where of course one uses $\fprimeos$ for local tomography acquisition (LTA) and
$\fprimeps$ for sinogram oriented acquisition (SOA).

The numerical studies in this work, which involve the simulation of 
data acquisition and reconstruction using both LTA and SOA, were performed
using a Python package we developed called ``Tomosim,'' which has been
made freely available on Github (https://github.com/mdw771/tomosim).
The charcoal tomographic dataset has been made available on TomoBank
\cite{decarlo_mst_2018} with a sample ID of 00078.

\subsection{Radiation dose calculation}

The differential energy deposition $dE$ within an infinitesimal depth 
$dt$ is formulated from the Lambert-Beer law as
\begin{equation}
	\frac{dE}{dt} = \Big|\bar{n}E_0\frac{dI}{dt}\Big|
	\label{eqn:diff_energy_didx}
\end{equation}
where $\bar{n}$ is the average number of incident photons per voxel,
and $E_0$ is the photon energy. 
The Lambert-Beer law gives $\mu_{\boldsymbol{r}}(t)$, the x-ray 
LAC of the sample as a function of 
penetration depth $t$ along the current transmission direction 
$\boldsymbol{r}$, as $\mu_{\boldsymbol{r}}(t) = -[1/I(t)](dI/dt)$.
To simplify our later computation with this term included in an
integral with regards to $t$, we approximate the quantity 
$I(t)$ in the factor prior to $dI/dt$ as $I(t) \approx I(L/2) = 
\exp(-\bar{\mu}L/2)$, where $\bar{\mu}$ is the mean LAC of the
specimen. Equation~\ref{eqn:diff_energy_didx} then becomes
\begin{equation}
	\frac{dE}{dt} = \bar{n}E_0\exp(-\bar{\mu}L/2)\mu_{\boldsymbol{r}}(t).
	\label{eqn:diff_energy}
\end{equation}
Again, the term $\exp(-\bar{\mu}L/2)$ represents the
beam attenuation factor at the center of the object, but it can also
be used to approximate the average normalized beam intensity ``seen''
by an arbitrary voxel of the object in one viewing direction. Accordingly,
we also replace $\mu_{\boldsymbol{r}}(t)$ in Eq.~\ref{eqn:diff_energy} 
with a constant value of $\bar{\mu}$. This approximation is valid
as long as the LAC of the object varies slowly. By integrating both
sides over the voxel size $\Delta$, we obtain the energy absorbed
by this voxel as
\begin{equation}
	E = \bar{n}E_0\exp(-\bar{\mu}L/2)\bar{\mu}\Delta.
	\label{eqn:diff_energy2}
\end{equation}
Then, the radiation dose received by this $j$-th voxel per (\ang{180}) scan 
is given by
\begin{equation}
	D_{s,j} = \frac{N_{\theta}\bar{n}E_0\exp(-\bar{\mu}L/2)\bar{\mu}}{\rho\Delta^{2}}
	\label{eqn:energy_radon}
\end{equation}
where $N_{\theta}$ is the number of projection angles, and
$\rho$ is the object density.
The subscript $s$ in $D_{s,j}$ denotes the $s$-th scan. Again, for SOA,
a total of $\nscanps$ scans are needed, while for LTA the number is on the
order of $\nscanos^2$. 
Based on this, one can estimate the total radiation dose received by
the sample by summing up the number of occasions of being exposed to
the beam over all voxels ($j$) and all scans ($s$). This is compactly expressed
as
\begin{eqnarray}
	D &=& \sum_s\sum_j D_{s,j} \nonumber \\
	&\propto & \sum_s \epsilon\Omega_s
	\label{eqn:total_dose}
\end{eqnarray}
where $\Omega_s$ is the total area in sinogram domain
that is sampled in one scan (which is equal to the width in pixels of a field of view
multiplied by the number of projection angles), and $\epsilon$ is the fraction of
pixels where the sample is present (\textit{i.e.}, pixels that are not
purely air).

\subsection{Experimental data acquisition and registration}

For an experimental tests on sinogram oriented acquisition (SOA), we
used experimental data collected using 25 keV X rays at beamline 32-ID
of the Advanced Photon Source at the Argonne National Laboratory. 
The specimen is a truncated charcoal sample
with a diameter of $d=4$ mm, whereas the imaging system field of view
was $f=1920\times 0.6$ $\mu$m=1.12 mm.  With $\gammaps=0.9$,
this yields a reduced field of view of $\fprimeps=1.04$ mm so that
$\nscanps=4$ and $T=0.26$.  Registration of the sinograms was done using phase
correlation, which can be formulated as
\begin{equation}
	\boldsymbol{c} = \argmax_{x\in\mathbb{R}^2} 
	    \mathcal{F}^{-1}
	    \Bigg[\frac{\mathcal{F}[I_a(\boldsymbol{x})] \mathcal{F}[I^*_b(\boldsymbol{x})]}
	               {|\mathcal{F}[I_a(\boldsymbol{x})] \mathcal{F}[I^*_b(\boldsymbol{x})]|}
	    \Bigg](\boldsymbol{x}).
\end{equation}
This method is reliable when a large number of high-contrast features are
present in the overlapping region of both images $I_a$ and $I_b$, and when
noises are not heavily present. In practice, photon flux ($\bar{n}$ in
Eq.~\ref{eqn:energy_radon}) sometimes needs to be reduced in order to
lower the radiation dose imposed on the sample. This can lead to higher photon
noise that challenges image registration. 

\section{Results and discussion}

\subsection{Comparison on dose-efficiency}

As one can easily see from Fig.~\ref{fig:acquisition}, object
stitching (LTA) requires a larger number of scans
than sinogram oriented acquisition (SOA) by a factor of about $1/T$. Because
much of the illumination of one scan goes into out-of-local-tomogram
regions in local tomography acquisition, this also means that the object is
exposed to a higher radiation dose.  
In Eq.~\ref{eqn:total_dose}, the total radiation 
dose of an experiment
is shown to be approximately proportional to the area of non-air regions
sampled in the sinogram, given that the 
sample does not contain large fluctuations 
in absorption coefficient. In this equation, $\Omega_s$
itself is also an interesting quantity to
investigate. The sum of the areas of all $\Omega_s$ regions in the
sinogram, which also includes those ``air'' pixels, provides an 
intuitive measurement of the acquired data size, which is jointly 
determined by the actual field of view, the
number of scans $s$, and the number of projection angles $N_{\theta}$. For a given
experimental configuration, this summed area is denoted by $A$.
A lower $A$ means
that the sample can be entirely imaged and reconstructed with a
smaller data size (\emph{i.e.}, less disk space is needed to store a complete acquisition), 
which is desirable in the case
where only limited storage and computing resources are available. 

\begin{figure}
	\centerline{\includegraphics[width=0.95\textwidth]{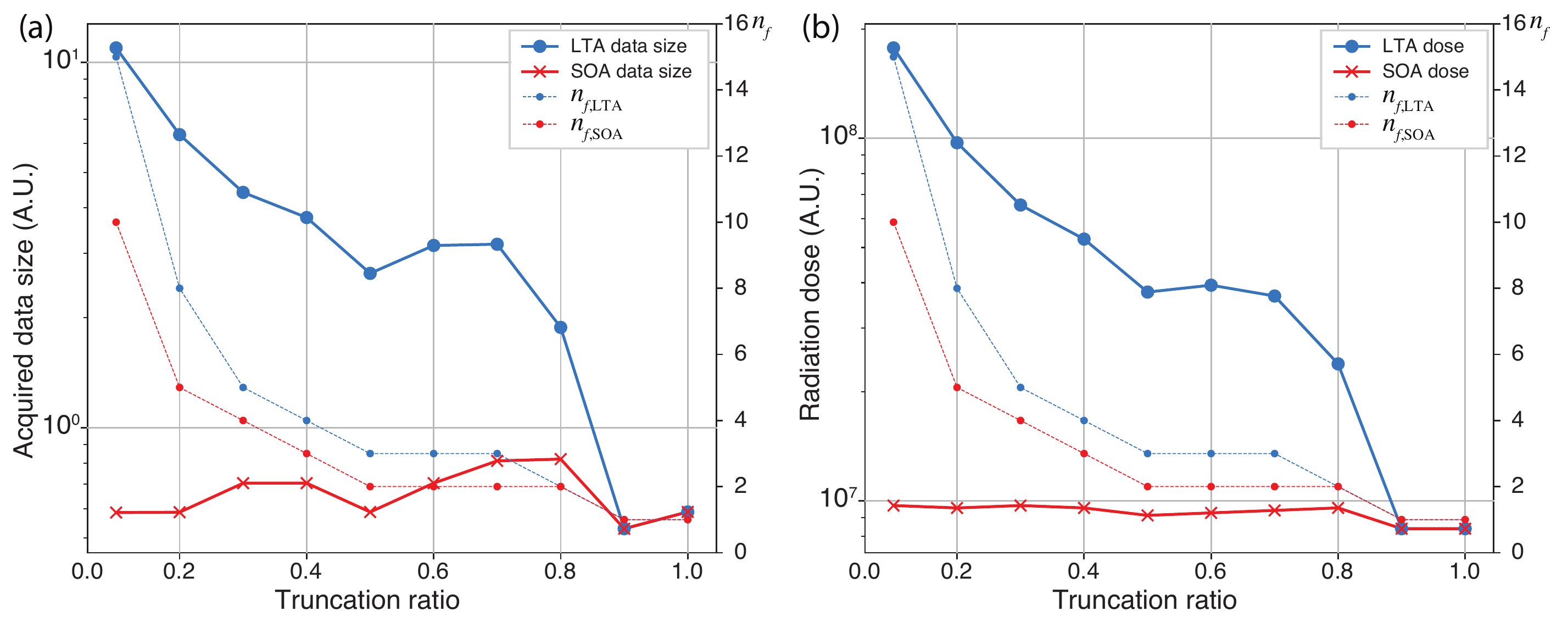}}
	\caption{Acquired data size (a) and radiation dose (b) as
          a function of the truncation ratio $T$ of
          Eq.~\ref{eqn:truncation_ratio} for both local tomography acquisition
          (LTA) and sinogram oriented acquisition (SOA).  In each subplot, the
          variation of $\nscanps$ and $\nscanos$ is also shown. The
          figure indicates that the acquired data size and radiation
          dose do not necessarily decrease with increasing truncation
          ratio; rather, both quantities are associated with the
          arrangement of scans in an actual experiment. These results
        were calculated for fixed values of $\gammaps=0.85$ and
        $\gammaos=0.85$ as discussed in the text.}
	\label{fig:eff_ratio}
\end{figure}

With these quantities defined, Fig.~\ref{fig:eff_ratio}(a) shows the
results for acquired data size $A_\textup{SOA}$ and $A_\textup{LTA}$
as a function of truncation ratio $T$ , while
Fig.~\ref{fig:eff_ratio}(b) shows $D_\textup{SOA}$ and
$D_\textup{LTA}$.  The dashed lines in each plot show the variation of
$\nscanps$ and $\nscanos$.  Note that $\nscanos$ is the number of
scans along one side of the object, so that $\nscanos^{2}$ scans are
required for local tomography acquisition (LTA). When examining this
figure, it has to be noted that no matter what $T$ is, the values of
$\gammaps$ and $\gammaos$ are fixed. This means that area covered by
all scans in either SOA or LTA might be larger than the sample. In
such cases, we allow acquisition to extend beyond the right side of
the sample for SOA; for LTA, the exceeded margins are on the right and
and bottom sides of the sample. The ``overflow'' of acquisition does
not substantially affect $D$, but can increase $A$. It can be seen in
Fig.~\ref{fig:eff_ratio}(a) that $A$ is not a monotonic function of
$T$, although it does show an overall decreasing trend with increasing
$T$. For example, when $T$ grows from 0.5 to 0.7, $A_\textup{LTA}$
increases while $\nscanos$ is unchanged. This is explained by the
larger ``overflow'' of scanned field beyond the actual object. In
contrast, the increase in $T$ that does not cause a reduction of
$\nscan$ only results in a small cost of $D$ due to the increase of
overlapping areas required between adjacent scans.  However, the
overall observation is still that SOA is both more data-efficient and
dose-efficient than LTA in general. The figures indicate that
no matter which method is used, a higher $T$ does not necessarily
imply better data efficiency in the case of $f < L$.  One should
thus carefully choose the camera to use in order to optimize the
experiment in terms of both data size and radiation dose.

%Note that in this case, a truncation ratio of $T = 1$
%does not necessarily imply a single FOV for both methods, since the
%FOV is expanded as needed to guarantee the full coverage of the whole
%object for small $\nscan$. Given the value of $\gamma$, the case where
%$\nscan = 1$ corresponds to a truncation ratio of $T= 1.52$. 
%In addition, since the object is circular in shape, LTA scans that fall 
%on the corner of the object array and do not have any overlap with the
%non-vacuum regions are discarded, so that they do not induce unnecessary
%exposure to the sample. It can
%be seen that LTA possesses advantage in either sampling efficiency
%or dose efficiency over SOA only in the case where the truncation ratio
%of LTA is several times higher than that of SOA. With the same detector,
%SOA always deposits less radiation energy. For a homogeneous sample
%with roughly equal extension along both directions on the rotation plane,
%the radiation dose of LTA is approximately higher than that of SOA
%by a factor of $\nscan$, though deviation from this tule-of-thumb can be
%seen for small truncation ratio (or large $\nscan$), in which case
%LTA scans near the edge of the object leave the bulk part of the
%specimen out of the x-ray beam for the majority of the rotation cycle,
%resulting in a smaller radiation dose contribution from these positions.

\subsection{Comparison on reconstruction artifacts}

\begin{figure}
	\centerline{\includegraphics[width=0.9\textwidth]{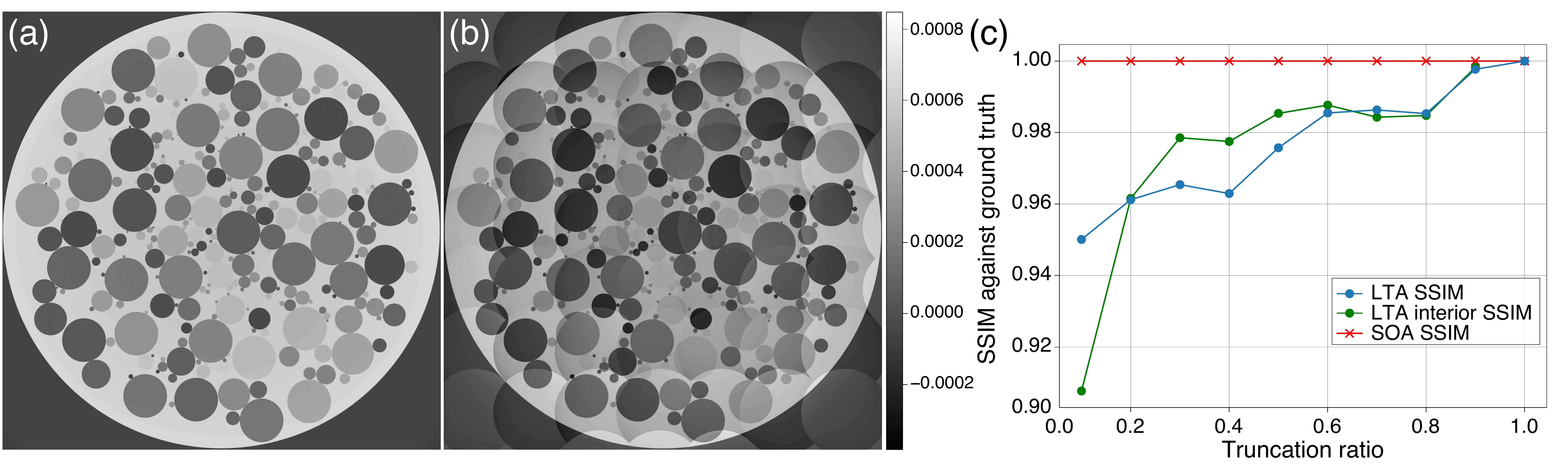}}
	\caption{Comparison of image quality between local tomography
          acquisition (LTA) and sinogram oriented acquisition
          (SOA). This comparison is made using the Structural
          SIMilarity index (SSIM) with regards to the ground truth
          image, against reconstructions for SOA (a) and LTA (b). When
          noise is not a factor, and registration errors are
          negligible, the SOA result is identical to the ground
          truth. Also shown (c) is the SSIM as a function of
          truncation ratio $T$.}
	\label{fig:recon_ssim}
\end{figure}

While both LTA and SOA are subject to photon noise during measurement,
other types of artifacts can also participate in determining the 
reconstruction quality.
The sources of noise and artifacts in the final reconstructions for SOA
are straightforward to understand. In particular, when the intensity
of adjacent projection tiles differs, ring artifacts can be seen in
the reconstruction if the sinograms are not properly blended where
they overlap. For LTA, reconstruction quality is mainly affected by
three factors other than noise in the raw data. First, since the
illuminating beam at different scan positions and illumination angles
arrive at the object region with varying transmission through
out-of-object-field features, the overall intensity of the
reconstruction disk can vary between neighboring tiles. 
This issue can be mitigated by gradient-based image blending techniques
such as Poisson blending \cite{perez_acm_2003}, but they are usually
time-consuming and are not appropriate when the number of tiles is
large. Second, a
bowl-shaped intensity profile across an individual reconstruction disk
is often observed in ROI tomography, in which case the pixel
intensities near the edge of the reconstruction disk are shifted
higher. This can be alleviated by padding the partial sinograms on its
left and right sides (along the spatial axis) by the edge values
\cite{kyrieleis_jmicro_2010}. In our case, the sinograms were padded
by twice their length on each side, but this did not completely
eliminate distortion in the intensity profile.  Finally, each
projection image collected inevitably contains information of the
portion object lying outside of the ROI, which, at least to some minor
extent, violates the Fourier slice theorem \cite{kak_2012}. When the
truncation ratio is not too low, one can use this excessive information
to slightly expand the field-of-view by padding both sides of the sinogram
with its edge values; however, streak artifacts will be heavily present
in the area out of the scanned disk in the case of a small truncation
ratio \cite{dasilva_optexp_2018}. 
In addition, ideally,
one would also seek to satisfy the Crowther criterion
\cite{crowther_prsa_1970} on the required number of rotation angles
based on the entire object size rather than the size of the local
tomography region of interest.  One can thus expect aliasing artifacts
especially for low truncation ratios.

To quantify the reconstruction quality, we used Structural SIMilarity
(SSIM; \cite{wang_ieeetip_2004}) as a metric for the fidelity of the
stitched reconstruction images with regards to the ground truth image.
The SSIM allows us to independently examine the structural fidelity of
an image with regards to the reference by incorporating the 
inter-dependency of image pixels, especially when they are spatially close. 
These dependencies carry important information about the structure, so that 
it serves as an accurate and reliable tool for evaluation. The reconstruction
images were obtained by applying filtered backprojection (FBP)
algorithm to the full-object sinogram. SSIM is defined as a product of
three terms that assess the similarity between two images $A$ and $B$
in different aspects. These include the luminance ($l$), the contrast
($c$), and the structure ($s$), defined by
\begin{eqnarray}
	l(A, B) &=& \frac{2\mu_A\mu_B + c_1}{\mu_A^2 + \mu_B^2 + c_1} \label{eqn:ssim_l} \\ 
	c(A, B) &=& \frac{2\sigma_A\sigma_B + c_2}{\sigma_A^2 + \sigma_B^2 + c_2} \label{eqn:ssim_c} \\
	s(A, B) &=& \frac{\sigma_{AB} + c_3}{\sigma_A\sigma_B + c_3} \label{eqn:ssim_s} 
\end{eqnarray}
where
\begin{eqnarray}
	c_1 &=& (k_1 L)^2 \\
	c_2 &=& (k_2 L)^2 \\
	c_3 &=& c_2 / 2 
\end{eqnarray}
with typical values of $k_1$ and $k_2$ set to 0.01 and 0.03, and $L$ being the
dynamic range of the grayscale images. In Eqs.~\ref{eqn:ssim_l} to \ref{eqn:ssim_s},
$\mu_i$ and $\sigma_i$ represent the mean and standard deviation of image $i$ ($i = A$
or $B$), and $\sigma_{AB}$ is the covariance of image $A$ and $B$ \cite{wang_ieeetip_2004}.
While it is common to calculate the SSIM as the product of all three terms, we set
$l(A, B) = 1$ here in order to exclude the overall intensity shifting and scaling.
Thus for all quality evaluations in this work, we have
\begin{equation}
	\mbox{SSIM}(A, B) = c(A, B)\cdot s(A, B).
\end{equation}
Figs.~\ref{fig:recon_ssim}(a) and (b) respectively show the stitched reconstructions
obtained from SOA and LTA with $T = 0.2$. If the beam brightness
is sufficiently high and stable, then noise and intensity variations
between adjacent tiles can be neglected. In this case, the stitched sinogram in SOA
is not affected by other systematic artifacts, and is identical to the full-object
sinogram. However, the stitched reconstruction in LTA is
affected by intensity variations and bowl-profile artifacts, even though the partial
sinograms were padded before reconstruction and only the inner $\gammaos = 0.85$ of
the reconstructed ROIs were used. Fig.~\ref{fig:recon_ssim}
plots the SSIM of the reconstructed porous disk (vacuum portions at the corners are not
included) with regards to the ground truth for both two approaches. As can be seen,
the quality of the SOA reconstruction is in principle not affected by the truncation
ratio. In contrast, an overall reduction in SSIM with decreasing truncation ratio 
is seen for LTA. 
%The fluctuations for lower truncation ratios are expected since as the
%FOV approaches the size of the pores, more statistical undulations arise.
In order to examine how the truncation ratio $T$ influences the reconstruction quality of
an individual reconstruction disk in LTA, we also computed the mean SSIM of the inner
portions in all reconstructed ROIs that are far from the boundaries
and termed it the ``LTA interior SSIM'' in Fig.~\ref{fig:recon_ssim}.
In this way, the influence of the bowl-profile artifacts 
can be excluded. As in 
Fig.~\ref{fig:recon_ssim}, this SSIM also drops with diminishing $T$. This indicates
that in addition to boundary artifacts, a low truncation ratio also deteriorates
the intrinsic reconstruction quality of the ROI, which is mainly in the form of
noise caused by out-of-ROI information. 

\subsection{Comparison on image registration feasibility}

Registration refers to the processing of finding the relative
positional offset between adjacent tiles in mosaic tomography. For SOA,
registration is done in projection domain before merging the partial
datasets. LTA, on the other hand, involves registration on
reconstructed images. The large number of tiles in mosaic tomography
poses huge difficulties for manual registration, and automatic
registration methods are usually employed. Phase correlation (PC) is
the most popular registration algorithm, where the offset vector
$\vec{c}$ between two images $I_a$ and $I_b$ is determined by
\begin{equation}
	\vec{c} = 
	    \argmax\Bigg(\mathcal{F}^{-1}
	    \Bigg[\frac{\mathcal{F}[I_a(\vec{x})] \mathcal{F}[I^*_b(\vec{x})]}
	               {|\mathcal{F}[I_a(\vec{x})] \mathcal{F}[I^*_b(\vec{x})]|}
	    \Bigg](\vec{x})\Bigg).
  \label{eqn:shift_vector}
\end{equation}
In Eq.~\ref{eqn:shift_vector}, $\mathcal{F}$ is the Fourier transform operator, 
and $I^*_i$ represents the complex conjugate of $I_i$. 

\begin{figure}
	\centerline{\includegraphics[width=0.7\textwidth]{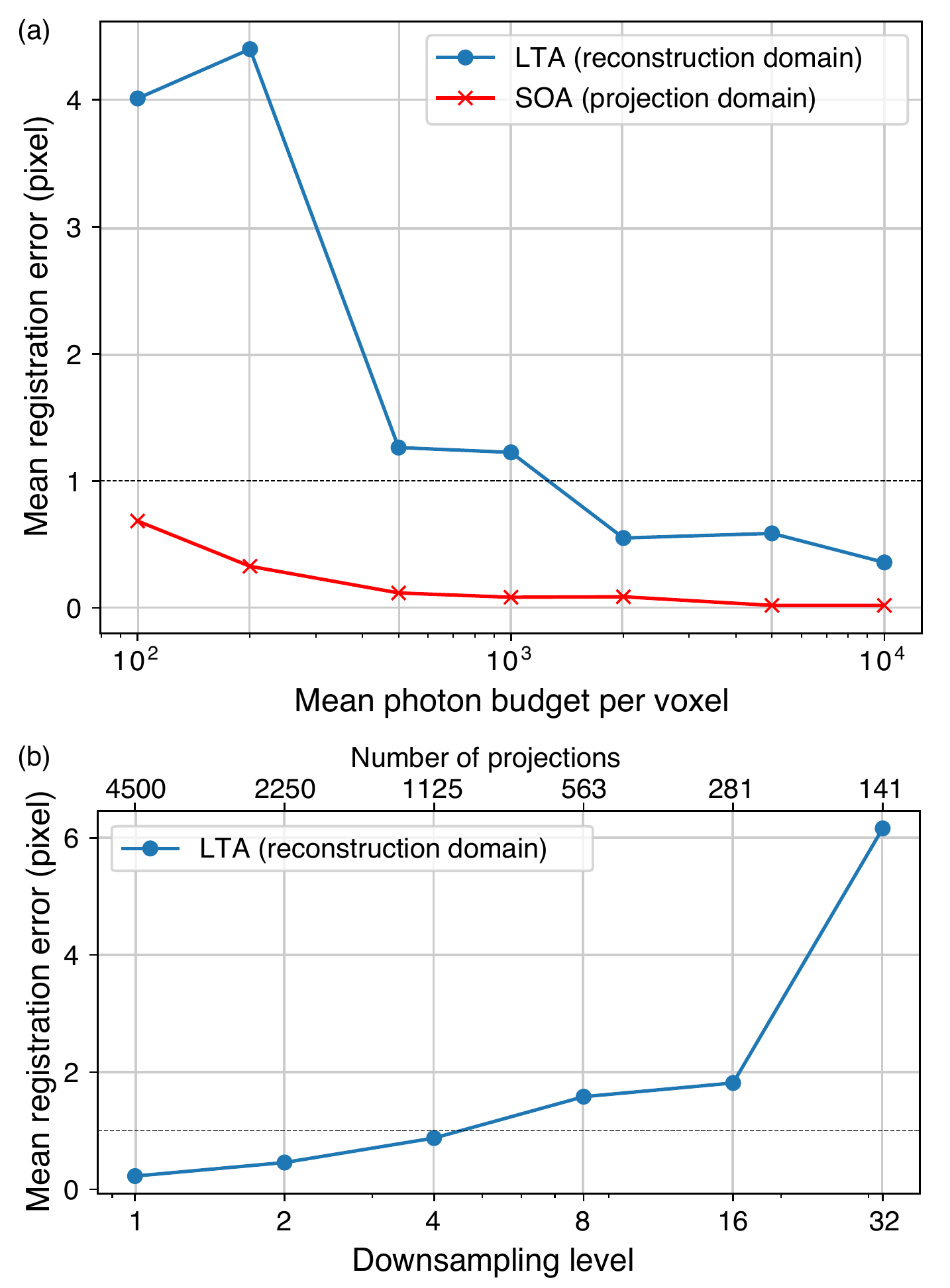}}
	\caption{Mean registration error plotted against (a) the
          average photon budget per voxel for both SOA and LTA, and (b)
          the downsampling level in projection angles for LTA.  The
          plot indicates that the accuracy of phase correlation
          degrades when projection images become more noisy due to
          lower number of incident photons. For this particular
          sample, registration in reconstruction domain for LTA
          requires a higher incident photon flux in order to give
          reliable registration results. In addition, a reduction in
          projection angles also causes a significant deterioration in
          registration accuracy for LTA.}
	\label{fig:error_phmult}
\end{figure}

The transmission radiographs for specimens that are thick and not
entirely periodic generally do not exhibit a good number of
distinguishable fine features, because the features tend to entangle
and blend into each other when they are superposed along the beam
path. However, this does not imply that SOA projections are
intrinsically harder objects for registration compared to LTA
reconstructions.  Although it is conceptually plausible that more
high-frequency features arise in reconstructed images, we should
notice that phase correlation is a technique that is susceptible to
noise. When data are collected with low photon flux, Poisson noise is
more pronounced, and tomographic reconstructions based on the Fourier
slice theorem can be more heavily affected by noise due to the
amplification of high-frequency artifacts by the ramp filter
\cite{kak_2012} (though this issue can be mitigated by adding a Weiner filter).
To investigate the photon noise sensitivity of alignment, we carried
out the following numerical study.
We created a
projection panorama of our whole charcoal specimen, and extracted a row
of $1024\times 1024$ pixel tiles from it with a constant interval of 850
pixels. As the projection panorama was normalized using the dark field
and white field data, all tiles extracted contain pixels with values
ranging between 0 and 1. Here we denote the image by $I$.  We then
define a scaling factor $n_{\rm ph}$ to represent a ``mean'' photon
count for each pixel. In other words, $n_{\rm ph}$ is the number of
photons incident on a pixel of the acquired radiograph.  Poisson
noises was subsequently applied to all tiles, using the probably
density function of
\begin{equation}
	f(k, n_{\rm ph}I) = \frac{(n_{\rm ph}I)^k e^{-n_{\rm ph}I}}{k!}
\end{equation}
where $k$ is the actual number of photon count. The noisy version of
the tiles were then pre-processed by taking their negative logarithm, and
registered using phase correlation. For LTA, different levels of
Poisson noise were added to extracted partial sinograms, from which
reconstruction images were subsequently created and registered. The
field of view in this case is 1024 pixels.  Since data fidelity is
guaranteed only within a disk for an LTA reconstruction, we use a
smaller offset of 700 pixels in both the $x$ and $y$ directions in
order to compensate the smaller usable overlapping area.
Figure~\ref{fig:error_phmult}(a) compares the registration accuracy of
LTA and SOA over a range of photon budgets per pixel, which is the
total number of photons to be applied to a specimen voxel during the
experiment. Thus, all comparisons between LTA and SOA are based on the
condition that the total radiation doses are equal. The photon budget
is evenly distributed to all scans and $n_{\rm ph}$ is calculated
accordingly, in which case LTA will have a lower $n_{\rm ph}$ in a
single scan compared to SOA.  For our test data, the mean registration
error of SOA is always below 1, while LTA requires a photon budget of
about 2000 for the mean error to diminish into the sub-pixel level.

The number of projection angles can also impact the registration accuracy for LTA
since it is done in the reconstruction domain. In Fig.~\ref{fig:error_phmult}(b),
the mean registration error is plotted with regards to the level of downsampling
in the axis of projection angles. The original data involve 4500 projections, 
which were downsampled by factors of powers of 2. The result indicates that the
error starts to exceed the pixel-level boundary when the downsampling level
is larger than 4.

\begin{figure}
  \centerline{\includegraphics[width=0.47\textwidth]{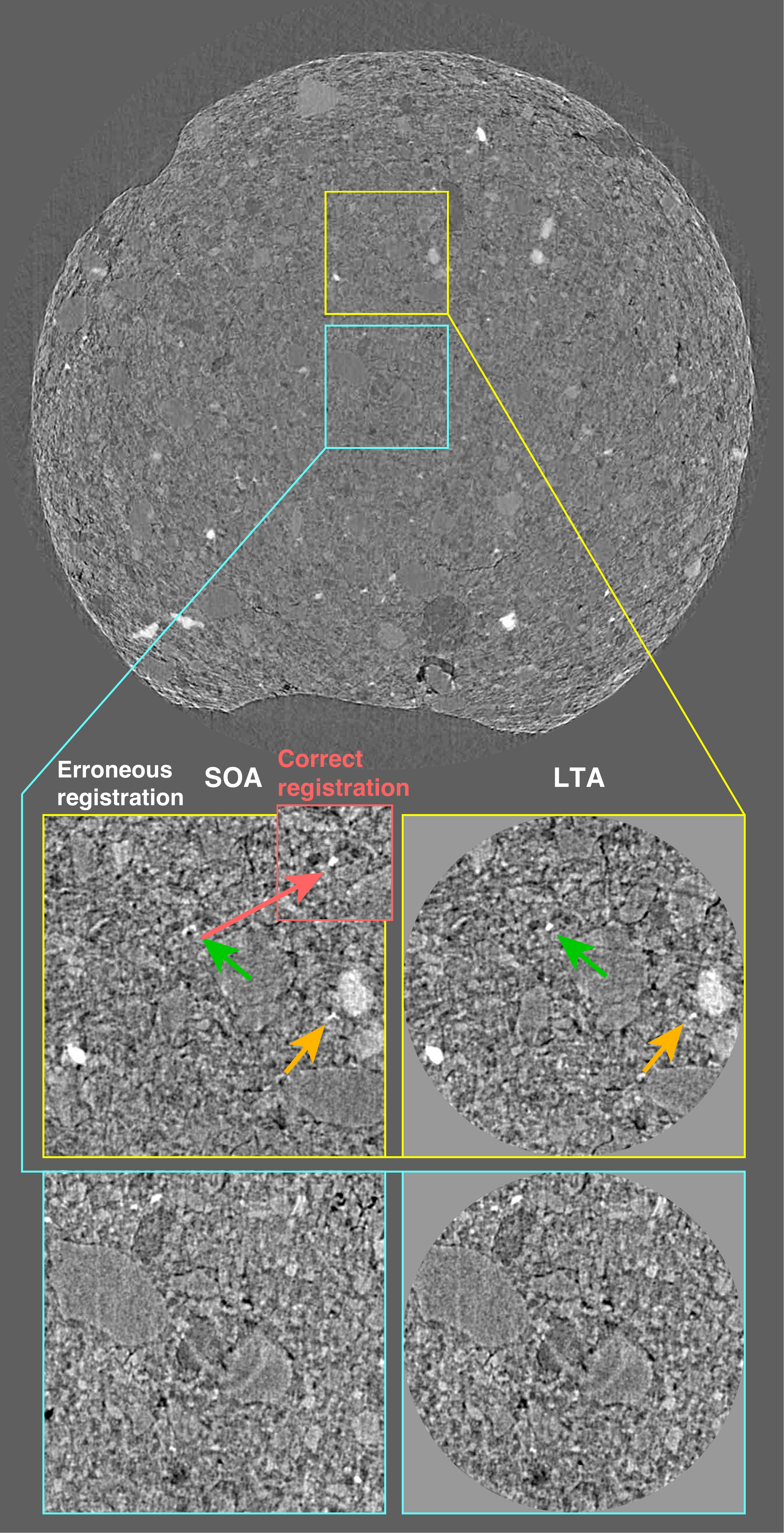}}
  \caption{Comparison of the effect of registration errors.  Shown
    here are sinogram oriented acquisition (SOA; left column in the
    grid) and local tomography acquisition (LTA; right column)
    reconstructions at an region-of-interest (first row) and the
    center (second row) of a slice in the charcoal sample. The SOA
    reconstruction was done by stitching 8 tiles extracted from the
    full sinogram.  Registration errors following a Gaussian
    distribution with a standard deviation of 4.0 were exerted to the
    tile positions before stitching.  The rotation centerfor SOA
    reconstruction was calibrated to optimize the quality around the
    object center. As a result, the central region of the charcoal
    reconstructed using both methods appears similar.  However, at
    around 1000 pixels above the object center, the SOA reconstruction
    shows severe distortion of dot-features (pointed by colored
    arrows) due to the deviation of its actual position from the
    rotation center inputted to the reconstruction routine. The inset
    in the SOA figure shows the appearance of one of the distorted
    features when the tiles are correctly registered. }
  \label{fig:register}
\end{figure}

A critical drawback of SOA is that registration errors are
accumulative, which means that deviations in the offset determined for
any pair of tiles can affect the quality of a large part or even the
entirety of the final reconstruction.  On the other hand, registration
errors in LTA involve multiple tiles intersecting on several sides,
giving less opportunity for alignment pathologies along one edge to
dominate global alignment.  For SOA, the accumulated registration
error throughout a row in the tile grid would cause the relative
center of rotation to deviate from the true value for tiles that are
far away from the rotation axis. Since the reconstruction of SOA takes
the registration results as an input, this can lead to off-center
distortions on small features at some locations of the full
reconstruction. To show this, we compare the reconstructions for a
part of the data collected from our charcoal sample. To simulate the
SOA result with induced error, we extracted 8 tiles from the full
sinogram with a fixed interval of 795 pixels. The registered positions
of all tiles were then deliberately adjusted by errors following a
Gaussian distribution with a standard deviation of 4, after which they
were stitched and reconstructed.  The center of rotation set for the
reconstructor was determined to optimize the reconstruction quality of
the central region of the sample.  The LTA results serving as a
reference were obtained by extracting partial sinograms from the full
sinogram, and then reconstructing them individually.  Using these
procedures, we show in Fig.~\ref{fig:register} a comparison of two
local regions of the reconstructions obtained using SOA and LTA,
respectively. One of these regions is exactly at the object center,
while the other one is around 1000 pixels above in the object slice
view. The positions of both regions are marked on the full
reconstruction slice. For the central region (shown in the second row
of the image grid in Fig.~\ref{fig:register}), the exhibited images
have nearly the same quality. However, for the off-center region, some
dot-shaped features extracted from the SOA reconstruction become
heavily distorted (as marked by the colored arrows in the SOA figure
on the first row of the grid). This indicates an erroneous
registration outcome for the tile contributing to this region, which
deviates the distance of the projections contained in this tile to the
rotation center away from the accurate value. When the tiles are
correctly registered, as shown in the inset of the SOA figure, the
distortion no longer exists.

%It should
%be pointed out, however, that the issue presented here is specific to
%registration methods working on the projection domain, and can be potentially
%bypassed by alternative registration strategies. For example, if the number
%of tiles in a row is not too large, one may register the tiles by reconstructing
%them individually, and align the resliced cross sections of each tile pair,
%where well resolved high-frequency features are available in a large number.

Local tomography acquisition (LTA) reconstructions are not globally
affected by registration errors. We further note that in addition to
this feature, LTA is advantageous compared to SOA in several other
aspects. For certain sample geometries, LTA can achieve better dose
efficiency than SOA by using more projection angles for highly
interesting regions of the sample while using fewer angles for the
rest.  Also, LTA allows one to flexibly select reconstruction methods
or parameters for different ROIs. For example, an ROI where features
lie in textured backgrounds can be reconstructed using Bayesian
methods with stronger sparsity regularization in order to suppress
background structures.

\section{Conclusion}

We have compared two methods for tomography of objects that extend
beyond the field of view of the illumination system and camera, based
on their radiation dose, reconstruction fidelity, and the presence of
registration artifacts. Sinogram oriented acquisition (SOA) gives
lower radiation dose, and it is also generally free of inter-tile
intensity variations, in-tile intensity ``bowl'' artifacts, and noise
induced by out-of-local-tomogram information. In addition, tile
registration is shown to be no harder than with Local tomography
acquisition (LTA), especially when the noise level is high. The major
drawback of SOA is that registration errors are accumulative and can
affect the entire reconstruction. Our present efforts are directed
towards providing more reliable registration algorithms in order to
improve the reconstruction quality of SOA for thick amorphous samples;
one approach that offers promise is iterative reprojection
\cite{dengler_ultramic_1989,latham_spie9967,gursoy_scirep_2017},
though it will be computationally demanding for large datasets.

\section{Acknowledgement}

This research used resources of the Advanced Photon Source and the
Argonne Leadership Computing Facility, which are U.S. Department of
Energy (DOE) Office of Science User Facilities operated for the DOE
Office of Science by Argonne National Laboratory under Contract
No. DE-AC02-06CH11357.  We thank the National Institute of Mental
Health, National Institutes of Health, for support under grant U01
MH109100.  We also thank Vincent De Andrade for his help with
acquiring the data on the charcoal sample shown in the paper.

%\section*{References}

%\section*{References}
\bibliographystyle{naturemag}
\bibliography{mybib}

\end{document}